\documentclass[reprint, aps]{revtex4-1}

\usepackage{amsmath}
\usepackage{amssymb}
\usepackage{mathtools}
\usepackage[usenames, dvipsnames]{color}
\usepackage{hyperref}

\begin{document}
\title{Photon Subtraction by Many-Body Decoherence}

\begin{abstract}
We experimentally and theoretically investigate the scattering of a photonic quantum field from another stored in a strongly interacting atomic Rydberg ensemble. Considering the many-body limit of this problem, we derive an exact solution to the scattering-induced spatial decoherence of multiple stored photons, allowing for a rigorous understanding of the underlying dissipative quantum dynamics.  Combined with our experiments, this analysis reveals a correlated coherence-protection process in which the scattering from one excitation can shield all others from spatial decoherence. We discuss how this effect can be used to manipulate light at the quantum level, providing a robust mechanism for single-photon subtraction, and experimentally demonstrate this capability. 
\end{abstract}

\author{C. R. Murray$^1$, I. Mirgorodskiy$^{2}$, C. Tresp$^{3}$, C. Braun$^{3}$, A. Paris-Mandoki$^{3}$, A. V. Gorshkov$^4$, S. Hofferberth$^{3}$ and T. Pohl$^1$}

\address{$^1$ Center for Quantum Optics and Quantum Matter, Department of Physics and Astronomy, Aarhus University, Ny Munkegade 120, DK 8000 Aarhus C, Denmark}
\address{$^2$ 5. Phys. Inst. and Center for Integrated Quantum Science and Technology, Universit\"at Stuttgart, Pfaffenwaldring 57, 70569 Stuttgart, Germany}
\address{$^3$ Department of Physics, Chemistry and Pharmacy, University of Southern Denmark, Campusvej 55, 5230 Odense-M, Denmark}
\address{$^4$ Joint Quantum Institute and Joint Center for Quantum Information and Computer Science, NIST/University of Maryland, College Park, Maryland 20742, USA}

\maketitle

Dissipation in quantum many-body systems can provide a powerful resource for realizing and harnessing a wide variety of complex emergent phenomena \cite{Diehl2008}. This notion has since enabled new concepts and strategies in dissipative quantum computation \cite{Verstraete2009}, simulation \cite{Barreiro2011} and many-body physics \cite{Diehl2011, Reiter2016}. Quantum optics systems present natural settings for such physics since they are intrinsically driven and dissipative in nature. Here, the interplay between coherent driving, photon propagation and dissipation can give rise to a broad range of nonequilibrium phenomena \cite{Ramos2014, Pichler2015}. In combination with strong optical nonlinearities at the quantum level \cite{Chang2014, Birnbaum2005, Hwang2009, Faez2014, Maser2016, OShea2013, Volz2014, Shomroni2014, Rosenblum2016, Tiecke2014, Javadi2015, Reiserer2013, Reiserer2014, Douglas2016, Peyronel2012, Firstenberg2013, Thompson2017}, this is now opening up a new frontier in strongly correlated nonequilibrium physics with photons \cite{Carusotto2009, Nissen2012, Hafezi2013, Maghrebi2015, Zeuthen2017}. In this direction, electromagnetically induced transparency (EIT) \cite{Fleischhauer2005} in atomic Rydberg ensembles \cite{Saffman2010} has emerged as one of the most promising approaches \cite{Friedler2005, Petrosyan2011, Murray2016, Hofferberth2016} for achieving strong, and often dissipative, photon-photon interactions.

The nonlinearity in such systems arises from the Rydberg blockade \cite{Lukin2001} that prevents EIT for nearby photons, yielding strong nonlinear dispersion \cite{Firstenberg2013} or dissipation \cite{Pritchard2010, Peyronel2012}. This mechanism has been successfully employed for few-body applications, such as all-optical switches \cite{Baur2014, Tiarks2014, Gorniaczyk2014, Gorniaczyk2016} and two-photon phase gates \cite{Tiarks2016}, where in both cases an initially stored gate photon controls the state of a subsequently passing source photon. On the other hand, a deeper understanding of many-body dynamics in these systems still presents an outstanding and formidable challenge to both theory and experiment. While the formation of three-body photon bound states has been studied \cite{Gullans2017} and reported \cite{Liang2017} very recently, the observational signatures for the transition to many-body behavior have remained elusive. 

In this work, we undertake such an extension of previous two-body applications \cite{Baur2014, Tiarks2014, Gorniaczyk2014, Gorniaczyk2016} to multiple gate and source photons. Our experiments performed in this many-body regime indeed reveal clear deviations from previous theories \cite{Li2015a, Murray2016a} for single gate-photon states. Remarkably, it is possible to derive a closed solution of the general many-body problem that accounts for the interplay of coherent photon propagation, strong atom-atom interactions and dissipative processes in an exact fashion. The new theory provides an excellent description of our experiments and reveals a correlated decoherence protection mechanism, where source photon scattering off one gate excitation shields all others behind it from spatial decoherence. Studying this backaction on the stored excitations, we show how it can be exploited to subtract a single photon from the retrieved gate field, and provide an experimental demonstration of this capability. In this way, the role of the source and gate fields are reversed, where the source field is now used to manipulate the stored gate field. 

\begin{figure*}
\begin{center}
\includegraphics[width=\textwidth]{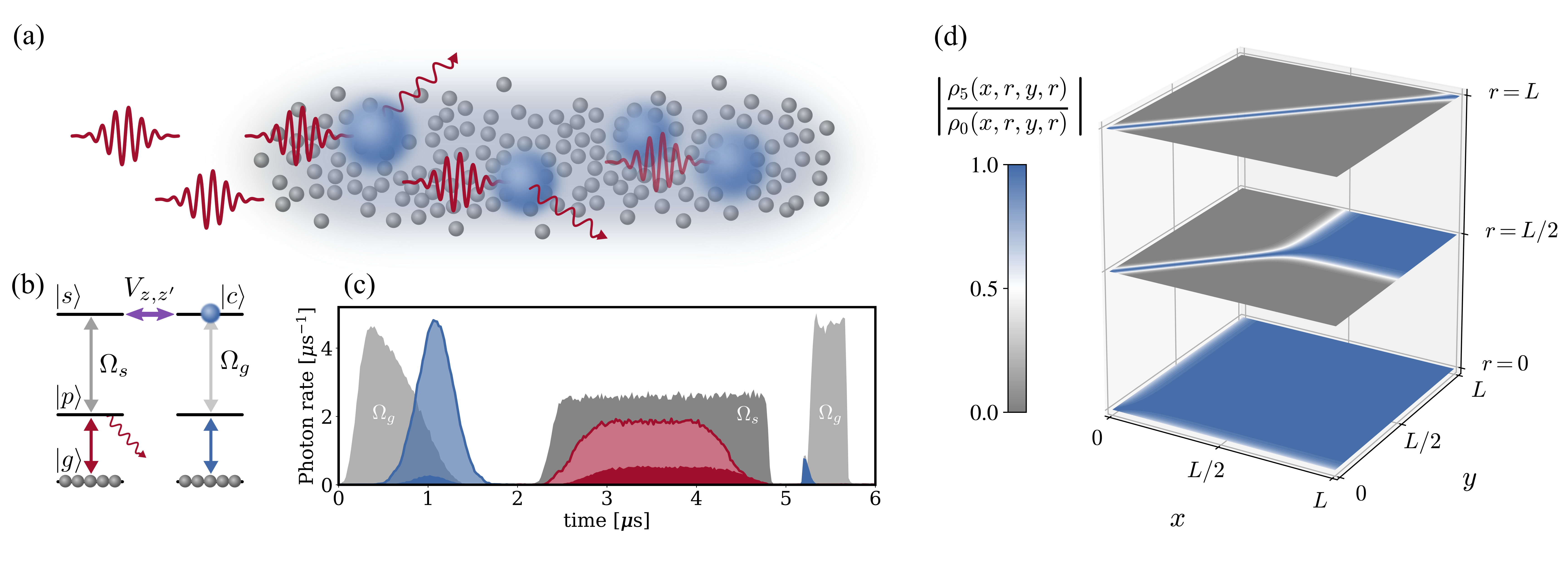}
\end{center}
\vspace{-5ex}
\caption{\label{fig: Figure 1} (a) Schematics of the basic setting in which Rydberg spin wave excitations (blue) stored in an ultracold gas interact with secondary Rydberg polaritons formed by propagating photons (red), whose interaction-induced scattering causes decoherence of the stored spin waves. The underlying level scheme through which the initial gate (blue) and secondary source (red) photons are coupled to their respective Rydberg states $|c\rangle$ and $|s\rangle$ is shown in panel (b). Panel (c) shows the experimental pulse sequence for a complete cycle of storage, interaction and retrieval stages. The blue and red curves show the gate and source field envelopes respectively, where the light  and dark traces indicate the incident and transmitted intensities. The gray curves show the control field envelopes (not to scale). Panel (d) shows the density matrix $\rho_{5}(x, r, y, r)$ of two stored gate excitations after scattering $5$ source photons and illustrates the correlated nature of the associated decoherence process.}
\end{figure*}

The basic idea and setup are illustrated in Fig. \ref{fig: Figure 1}(a-c). Initially, a multiphoton gate field is stored \cite{Gorshkov2007, Gorshkov2007a, Novikova2007} as a collective spin wave in the Rydberg state $|c\rangle$ of an atomic ensemble to yield a system of $n_g$ stored excitations. This is achieved via Rydberg EIT with a properly timed gate-photon pulse and control field with Rabi frequency $\Omega_g$ as shown in Fig. \ref{fig: Figure 1}(c). Subsequently, a second source field containing $n_s$ photons is sent through the medium under EIT conditions with a different Rydberg state $|s\rangle$. The strong van der Waals interaction between $|s\rangle$ and $|c\rangle$ results in a spatially dependant level shift $V_{z, z^{\prime}}=C_6 / |z - z^{\prime}|^6$ of $|s\rangle$, where $z$ and $z^{\prime}$ are the positions of $|s\rangle$ and $|c\rangle$ respectively. This exposes the propagating source photons to a dissipative two-level medium of extent $2z_b$ surrounding each gate excitation. Here $z_b$ denotes the blockade radius \cite{Gorshkov2011} within which the formation of a dark state polariton is blocked. The effective optical depth of this exposed medium is $\sim 4d_b$, where $2d_b$ is the optical depth per blockade radius. For large values of $d_b$ nearly all incoming source photons are scattered in the blockade region such that this setup can function as an efficient optical switch \cite{Baur2014, Tiarks2014, Gorniaczyk2014, Gorniaczyk2016}.

This scattering, however, does not leave the gate photons unaffected. Each source photon scattered off a blockade sphere carries information about the position of the Rydberg excitation that is causing the blockade \cite{Murray2016a}. The associated coherence loss from such projective spatial measurements typically leads to strong localization of the original spin wave state, thereby inhibiting its subsequent retrieval. 

Formulating the described system in second quantization, we introduce the bosonic operator $\hat{\mathcal{E}}^{\dagger}(z, t)$ for the creation of a source photon at position $z$ and time $t$, and similarly $\hat{P}^{\dagger}(z, t)$, $\hat{S}^{\dagger}(z, t)$, and $\hat{C}^{\dagger}(z, t)$ for the creation of collective atomic excitations in the states $|p\rangle$, $|s\rangle$ and $|c\rangle$, respectively [see Fig. \ref{fig: Figure 1}(b)]. To describe the many-body decoherence dynamics of the stored excitations, we define the operator $\hat{\rho}(\vec{x}_{n_g}, \vec{y}_{n_g}, t) = \prod_{i=1}^{n_g} \hat{C}^{\dagger}(x_i, t) \prod_{i=1}^{n_g} \hat{C}(y_i, t)$ which characterizes the spatial coherence between different configurations $\vec{x}_{n_g}\equiv x_1, x_2,\cdots,x_{n_g}$ and $\vec{y}_{n_g}\equiv y_1, y_2,\cdots,y_{n_g}$ of the stored excitations. The dynamics of this operator is governed by the following equation of motion,
\begin{equation}\label{eq: Coherence equation of motion}
\begin{split}
\partial_t \hat{\rho}(\vec{x}_{n_g}, \vec{y}_{n_g}, t) = i & \int_{0}^{L} dz \left[ \sum_k V_{z, x_k} - \sum_k V_{z, y_k} \right] \\
& \times \hat{S}^{\dagger}(z, t) \hat{\rho}(\vec{x}_{n_g}, \vec{y}_{n_g}, t) \hat{S}(z, t).
\end{split}
\end{equation}
 Here, we assume low-intensity source and gate fields and neglect the source-source and gate-gate interactions. To calculate the spin wave decoherence predicted by Eq. (\ref{eq: Coherence equation of motion}), we start from the initial system state $| \Psi_{n_g, n_s}\rangle$ of $n_g$ stored gate excitations and $n_s$ incident source photons. The elements of the stored spin wave density matrix can then be defined according to $\rho_{n_s}(\vec{x}_{n_g}, \vec{y}_{n_g}, t) = \langle \Psi_{n_g, n_s}| \hat{\rho}(\vec{x}_{n_g}, \vec{y}_{n_g}, t) | \Psi_{n_g, n_s}\rangle$. Solving the dynamics of $\rho_{n_s}(\vec{x}_{n_g}, \vec{y}_{n_g}, t)$ according to Eq. (\ref{eq: Coherence equation of motion}) to zeroth order in the source field bandwidth, the final state of the stored gate excitations $\rho_{n_s}(\vec{x}_{n_g}, \vec{y}_{n_g}) = \rho_{n_s}(\vec{x}_{n_g}, \vec{y}_{n_g}, t\to\infty)$ after the passage of all source photons can be calculated as
\begin{equation}
\label{eq: Density matrix}
\rho_{n_s}(\vec{x}_{n_g}, \vec{y}_{n_g}) = \left[\Phi_{n_g}(\vec{x}_{n_g}, \vec{y}_{n_g}) \right]^{n_s} \rho_{0}(\vec{x}_{n_g}, \vec{y}_{n_g}),
\end{equation}
where $\rho_{0}(\vec{x}_{n_g}, \vec{y}_{n_g})$ is the initial state, and the quantity $\Phi_{n_g}(\vec{x}_{n_g}, \vec{y}_{n_g})$ is given by
\begin{equation}\label{eq: Phi}
\begin{split}
& \Phi_{n_g}(\vec{x}_{n_g}, \vec{y}_{n_g}) = 1 + \frac{d_b}{z_b} \int_0^L dz \frac{\sum_k \mathcal{V}_{z, x_k} - \sum_k \mathcal{V}_{z, y_k}}{\left(i + \sum_k  \mathcal{V}_{z, x_k} \right)\left(i -  \sum_k  \mathcal{V}_{z, y_k} \right)} \\
& \times \exp \left( \frac{d_b}{z_b} \int_0^z dz^{\prime} \left[\frac{\sum_k  \mathcal{V}_{z^{\prime}, y_k}}{i - \sum_k  \mathcal{V}_{z^{\prime}, y_k} } - \frac{\sum_k  \mathcal{V}_{z^{\prime}, x_k}}{i + \sum_k  \mathcal{V}_{z^{\prime}, x_k} } \right] \right).
\end{split}
\end{equation}
where $\mathcal{V}_{z, z^{\prime}} = \gamma V_{z, z^{\prime}} / \Omega^2$ is the rescaled interaction potential, and $\gamma$ is the decay rate of $|p\rangle$. A detailed derivation of this expression is presented in appendix \ref{sec: S1}.

The emergence of correlated decoherence can be readily understood by considering  a dilute system of gate excitations, where the contribution from spatial configurations with overlapping blockade radii can be neglected. Initially, the incoming source photons interact with the first gate excitation located closest to the incident medium boundary. As described above, the associated projective measurement of its position drastically degrades its retrieval. However, in the strong scattering limit, it also causes near complete extinction of the source field such that all subsequent gate excitations are shielded from photon scattering, leaving their spatial coherence unaffected. 

To reveal this effect from our solution, Eq. (\ref{eq: Density matrix}), consider the simplest situation of two gate excitations, now stored in the same spatial mode. The quantity $\rho_{n_s}(x, r, y, r)$ in this case characterizes how the local density component of one gate excitation, at a position $r$, affects the spatial coherence between $x$ and $y$ of the other excitation. In  Fig. \ref{fig: Figure 1}(d), we plot $\rho_{n_s}(x, r, y, r)$ for various values of $r$. Indeed, one finds that source photon scattering leads to almost complete decoherence, rendering $\rho_{n_s}(x, r, y, r)$ largely diagonal for $x, y < r$. For $x, y > r$, on the other hand, the coherence of one gate excitation with respect to $x$ and $y$ is preserved by scattering from the other excitation at position $r$. 

We can gain further insight into the decoherence dynamics for multiple gate excitations in the limit of $d_b\gg1$. In this case, the quantity $\Phi_{n_g}(\vec{x}_{n_g}, \vec{y}_{n_g})$ characterizing the final density matrix in Eq. (\ref{eq: Density matrix}) reduces to 
\begin{equation}
\Phi_{n_g}(\vec{x}_{n_g}, \vec{y}_{n_g}) \overset{d_b \to \infty}{\approx} \Phi_{1}(x_{\rm min}, y_{\rm min}),
\end{equation}
as shown in appendix \ref{sec: S2}, where $x_{\rm min} = \min\left\{ \vec{x}_{n_g} \right\}$ and $y_{\rm min} = \min\left\{ \vec{y}_{n_g} \right\}$ are the coherence coordinates of the first excitation. This result indeed shows that only the first excitation participates in the scattering dynamics. Since $\Phi_1(x_{\rm min}, y_{\rm min} \neq x_{\rm min}) = 0$ for $d_b \to \infty$, this explicitly shows that the coherence of this first excitation is vanishing. At the same time, it demonstrates that the photon scattering from its local density preserves the coherence of all other excitations, since $\Phi_1(x_{\rm min}, y_{\rm min} = x_{\rm min}) = 1$.

As described above, the efficiency of gate photon retrieval is directly affected by scattering induced spin wave decoherence. While this inhibits the retrieval of a single gate excitation \cite{Murray2016a}, the many-body decoherence protection between multiple gate excitations offers enhanced retrieval efficiencies, relative to the case of a single excitation. Here we derive a simplified description of gate photon retrieval from the full many-body density matrix $\rho_{n_s}(\vec{x}_{n_g}, \vec{y}_{n_g})$ in Eq. (\ref{eq: Density matrix}),  by assuming that scattering off one gate excitation leaves the mode shape, and thus retrieval efficiency, of all other excitations unaffected.  Considering coherent gate and source fields containing an average number of photons $\alpha_g$ and $\alpha_s$ respectively, we calculate the retrieval efficiency of each gate excitation sequentially from its reduced density matrix. The total retrieval efficiency $\eta$ can then be written as 
\begin{equation}\label{eq: Retrieval efficiency}
\eta = \eta_R \frac{e^{-\alpha_g}}{\alpha_g} \sum_{n_g = 1}^{\infty} \frac{(\alpha_g)^{n_g}}{n_g!} \sum_{k = 1}^{n_g} e^{- \alpha_s p (1 - p)^{(k - 1)} },
\end{equation}
as shown in appendix \ref{sec: S3}, where $p \approx 1 - \exp[-4d_b]$ is the source photon scattering probability per gate excitation, and $\eta_R$ denotes the retrieval efficiency in the absence of interactions between source and gate excitations.  The second summand in Eq. (\ref{eq: Retrieval efficiency}) is proportional to the probability of retrieving the $k^{\rm th}$ excitation in a given Fock state component of the stored field.  From this it is clear that in the strong scattering limit ($p\sim 1$), the retrieval of the first excitation ($k=1$) is suppressed, while the retrieval of all later excitations is ($k>1$) largely unaffected. The retrieval efficiency thus provides a well suited and accessible experimental probe of the many-body decoherence in the system.

\begin{figure}[h!]
\begin{center}
\includegraphics[width=0.95\columnwidth]{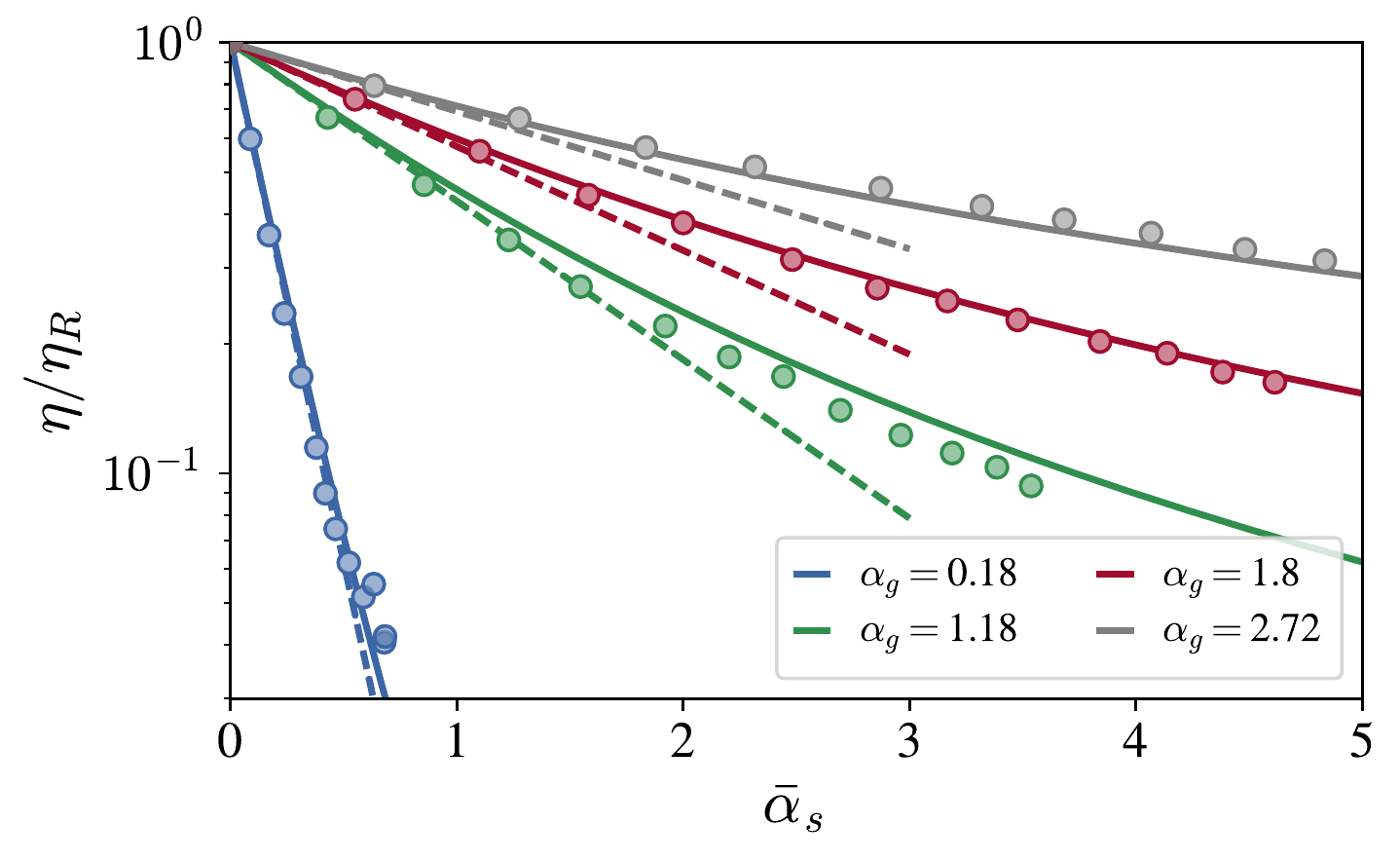}
\end{center}
\vspace{-5ex}
\caption{\label{fig: Retrieval efficiency} Normalized retrieval efficiency as a function of the number, $\bar{\alpha}_s$, of scattered source photons for different indicated numbers, $\alpha_g$, of stored gate excitations.  The theoretical prediction of Eq. (\ref{eq: Retrieval efficiency}) (solid lines) is fitted to the experimental data (dots) with a common scattering probability of $p = 0.5$ (errors bars showing SEM are smaller than the dots).   Retrieval efficiencies are on the order of $\eta_R\sim 0.2$ for all measured data.  The dashed lines indicate the expected scaling without decoherence protection.}
\end{figure}

Our experiments start by trapping $\sim9\times10^4$ $^{87}$Rb atoms into an optical dipole trap which yields a cigar shaped cloud at $4 \mu$K with $1/e$ radial and axial radii of $13 \mu$m and $42 \mu$m, respectively. All atoms are first optically pumped into the $|g\rangle=|5S_{1/2},F=2,m_F=2\rangle$ state. Gate photons are coupled to the Rydberg state $|c\rangle=|68S_{1/2},m_J=1/2\rangle$ via EIT by applying a weak $780$ nm probe field that drives the transition between $|g\rangle$ and the intermediate $|p\rangle=|5P_{3/2},F=3,m_F=3\rangle$ state. A strong counterpropagating $480$ nm control field drives the transition between $|p\rangle$ and $|c\rangle$ with a Rabi frequency $\Omega_g$ on two-photon resonance to establish EIT. We store gate photons in the cloud by turning off $\Omega_g$ while the gate photon pulse propagates through the cloud. The generated number of Rydberg excitations can be measured by standard field ionization detection from which we determine $\alpha_g$. Using a source photon pulse with an average number of $\alpha_s$ photons, we can probe the stored gate excitations optically by monitoring the source-photon transmission. In this case EIT is provided by another control laser that couples the intermediate state to the $|s\rangle=|66S_{1/2},m_J=1/2\rangle$ Rydberg state. Following their interaction with the source photons, the gate photons are read out by turning $\Omega_g$ back on after a total storage time of $4\mu$s. A typical complete pulse sequence is shown in Fig. \ref{fig: Figure 1}(c).

In Fig. \ref{fig: Retrieval efficiency} we show the retrieval efficiency as a function of the number, $\bar{\alpha}_s$, of gate-scattered source photons, which we determine from the transmission in the absence and presence of the gate excitations. If the photon-photon interactions would decohere all gate excitations, the retrieval efficiency would scale as $\eta_R \exp[- \bar{\alpha}_s / \alpha_g]$ which simply reflects the vacuum component of the source-photon pulse \cite{Murray2016a}. While this simple relation yields a good description for small $\alpha_g$ and $\bar{\alpha}_s$, we observe  significantly higher retrieval efficiencies for larger  photon numbers. Indeed, this can be traced back to the multiphoton protection mechanism introduced in this work, as further evidenced by the remarkably good agreement with the theoretical prediction of Eq. (\ref{eq: Retrieval efficiency}). 

\begin{figure}[h!]
\begin{center}
\includegraphics[width=0.95\columnwidth]{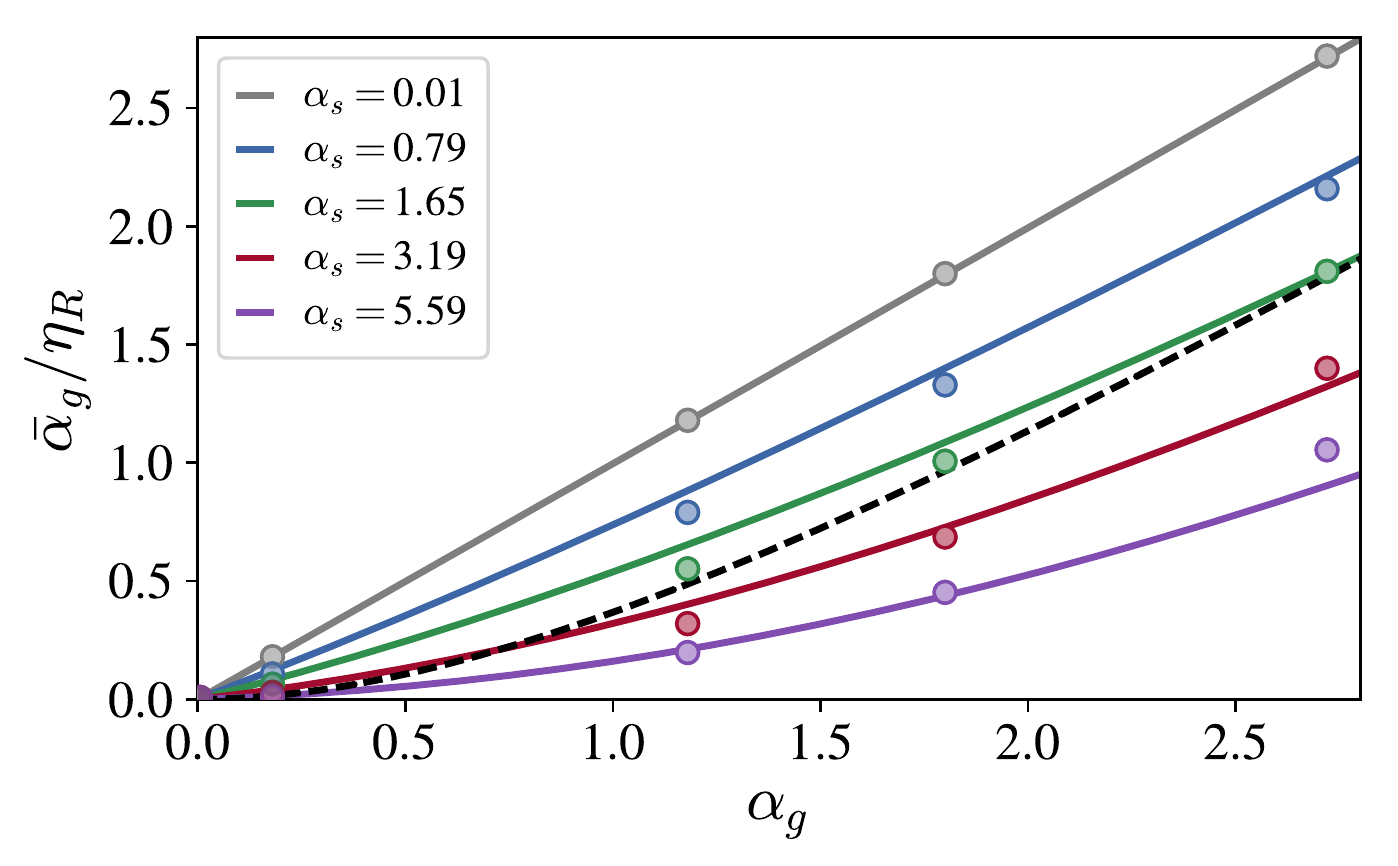}
\end{center}
\vspace{-5ex}
\caption{\label{fig: Subtractor} Number $\bar{\alpha}_g$ of retrieved gate photons as a function of the number $\alpha_g$ of initially stored excitations for different incident source photon numbers $\alpha_s$. The measurements (dots) agree well with the prediction of Eq. (\ref{eq: Retrieval efficiency}) (solid lines) for the same value of $p$ as in Fig. \ref{fig: Retrieval efficiency}. The black dashed line shows the prediction of Eq. (\ref{eq: Retrieval efficiency}) for $p=1$ and large source field intensity $\alpha_s \gg 1$, which are the ideal conditions for single-photon subtraction.}
\end{figure}

As the scattering probability approaches unity, only the first gate excitation participates in the decoherence dynamics. This in turn enables a robust mechanism for single-photon subtraction, since the inability to retrieve the decohered excitation effectively removes a single photon from the initial gate field upon retrieval. Fig. \ref{fig: Subtractor} shows the number $\bar{\alpha}_g$ of retrieved gate photons as a function of the number of stored gate excitations. Note that the number of subtracted photons can still exceed unity due to the imperfect scattering conditions, $p<1$, in the experiment. In this case, the first gate excitation does not completely extinguish the source field which can therefore decohere additional gate photons.
 For the source field intensities considered in Fig. \ref{fig: Subtractor}, the measured transmitted intensity is linear indicating that self-interactions between source photons have a negligible effect.

\begin{figure}[h!]
\begin{center}
\includegraphics[width=0.94\columnwidth]{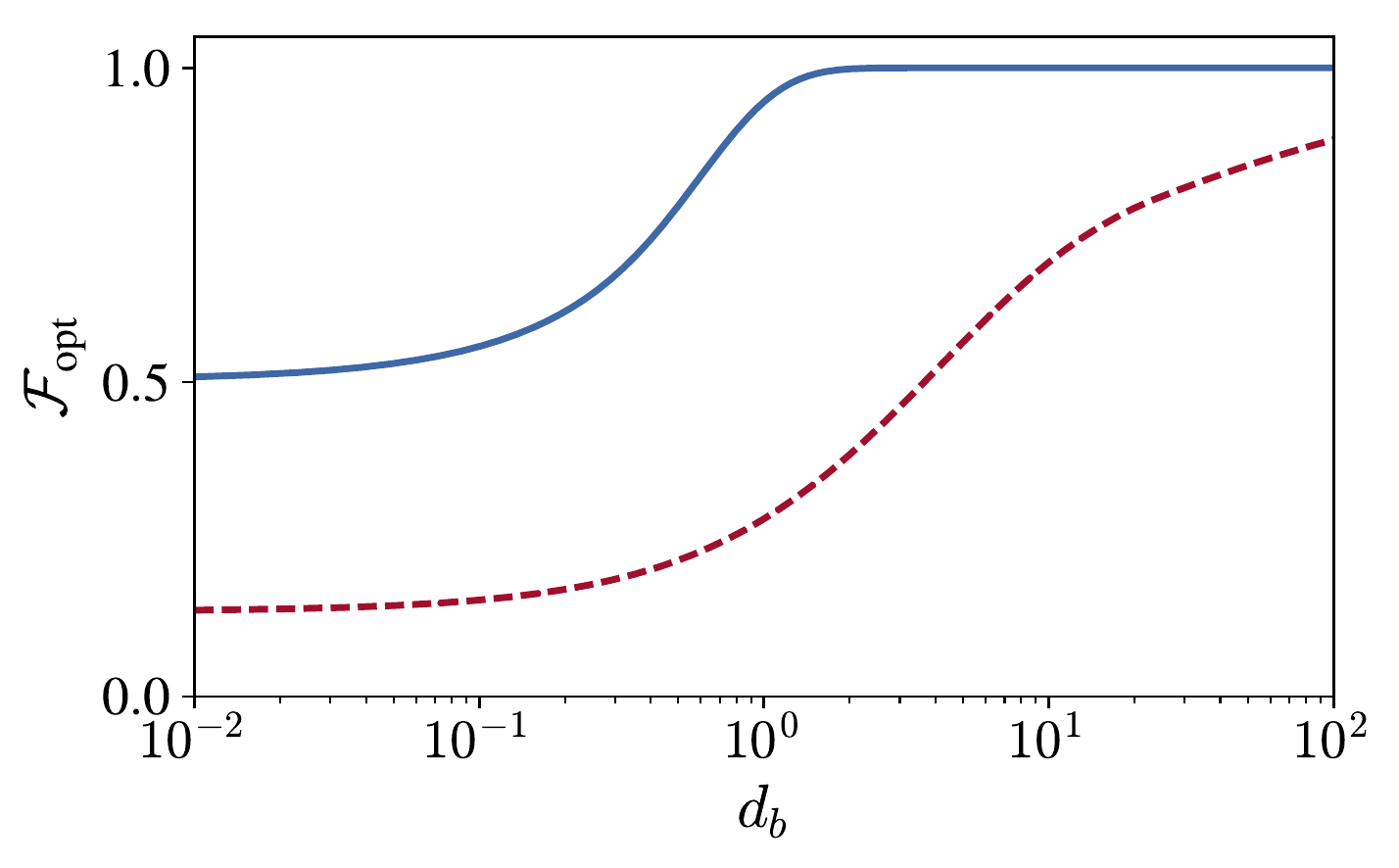}
\end{center}
\vspace{-4ex}
\caption{\label{fig: Efficiency} Single-photon subtraction efficiency, $\mathcal{F}_{\text{opt}}$, for a coherent gate field whith an average number of photons $\alpha_g = 2$.  The blue line shows the optimal efficiency of the current mechanism based on single-photon decoherence for perfect storage and retrieval, while the red dashed line shows the corresponding performance of photon subtraction by saturable absorption \cite{Tresp2016}.}
\end{figure}

To analyze the optimal operation of the photon subtractor, we define the probability $\mathcal{F}$ that exactly one photon is decohered by source photon scattering. Using the theory outlined above, we obtain
\begin{equation}\label{eq: Subtractor efficiency}
{\mathcal{F}} = e^{-\alpha_g} \left[ 1 + \sum_{n_g = 1}^{\infty} \frac{(\alpha_g)^{n_g}}{n_g!} P_1(n_g, \alpha_s) \right],
\end{equation}
as shown in appendix \ref{sec: S4}, where $P_1(n_g, \alpha_s)$ is the probability that the source field decoheres exactly one of the $n_g$ stored excitations in a given stored Fock state component. Upon maximizing Eq. (\ref{eq: Subtractor efficiency}) with respect to $\alpha_s$ we obtain the optimal subtraction efficiency $\mathcal{F}_{\text{opt}}$. We plot $\mathcal{F}_{\text{opt}}$ in Fig. \ref{fig: Efficiency}, and compare this to the corresponding performance of an alternative subtraction mechanism recently demonstrated in Ref. \cite{Tresp2016}. Such alternative schemes utilize quantum emitters whose absorption can be saturated by a single photon, e.g., through strong photon coupling to a single atom \cite{Rosenblum2016} or by exploiting the Rydberg blockade in atomic ensembles \cite{Honer2011, Tresp2016}. 

To draw this comparison, we have calculated the optimal subtraction efficiency of the approach demonstrated in Ref. \cite{Tresp2016}. The details of this calculation are outlined in appendix \ref{sec: S5}. Here one employs Rydberg state dephasing with a rate $\Gamma$ for efficient single-photon absorption with probability $p$. Working with a small ensemble, the produced Rydberg excitation then blocks the storage of subsequent photons and renders the medium largely transparent with a small residual absorption. While this strategy benefits from the growing single-photon absorption efficiency with increasing input power \cite{Tresp2016}, its fidelity is ultimately limited by the challenging requirement of maximizing $p$ at low residual photon absorption. In the present case, the scattering probability $p$ exponentially approaches unity with increasing $d_b$ which simultaneously enhances the protection of all other photons from decoherence, and thereby improves the overall subtractor performance.  Instead, the overall performance is limited by the finite storage and retrieval efficiency [see appendix \ref{sec: S4} for further details]. While the current experiment has not been optimized with respect to storage and retrieval, we note that recent measurements have reported combined efficiencies in excess of 95\% \cite{Hsiao2016}. Approaching this limit in Rydberg media would require longer clouds with higher optical depth and shorter storage times to minimize dephasing effects \cite{Mirgorodskiy2017}, combined with optimization of the storage and retrieval protocol \cite{Gorshkov2007, Gorshkov2007a, Murray2016a}.

In summary, we have investigated the dissipative quantum dynamics of multiple photons in a strongly interacting Rydberg ensemble. Considering the specific situation of stored Rydberg spin waves interacting with propagating Rydberg polaritons, we derived an exact solution to this general many-body problem, which reveals correlated spin wave dynamics and a mutual decoherence protection mechanism between multiple stored excitations. Our experiments clearly demonstrate this effect and suggest how it can be exploited to manipulate light at the quantum level. In particular we showed how the discovered effect can provide a robust mechanism for realizing a single-photon subtractor. Its current overall performance is limited by the efficiency for light storage and retrieval. Improving this capability and better understanding associated Rydberg-state effects \cite{Gaj2014, Schmidt2016, Liang2017, Camargo2017, Schmidt2017} will thus be central to future work, and is vital to a number of recent experiments \cite{Liang2017, Busche2017, Tiarks2016, Baur2014, Tiarks2014, Gorniaczyk2014, Gorniaczyk2016} based on light storage and subsequent photon interactions. Our measurements and developed theory of multiphoton decoherence effects provide valuable insights for such applications \cite{Tiarks2016, Baur2014, Tiarks2014, Gorniaczyk2014, Gorniaczyk2016} and future studies of strongly interacting Rydberg-polariton systems beyond the few photon limit.

We thank W. Li and I. Lesanovsky for useful discussions. This work is funded by the German Research Foundation (Emmy-Noether-grant HO 4787/1-1, GiRyd project HO 4787/1-3, GiRyd project PO 1622/1-1, SFB/TRR21 project C12), by the Ministry of Science, Research and the Arts of Baden-W\"{u}rttemberg (RiSC grant 33-7533.-30-10/37/1), by the EU (H2020-FETPROACT-2014 Grant No. 640378, RySQ), by ARL CDQI, NSF QIS, AFOSR, ARO, ARO MURI, and NSF PFC at JQI, and by the DNRF through a Niels Bohr Professorship.

\onecolumngrid
\appendix

\section{Gate excitation density matrix dynamics}
\label{sec: S1}
Here, we will outline the solution to the gate excitation density matrix given in Eq. (2) of the main text. To first describe the EIT dynamics of the source field, we introduce the bosonic operator $\hat{\mathcal{E}}^{\dagger}(z, t)$, which creates a source photon at position $z$ and time $t$, and similarly introduce the operators $\hat{P}^{\dagger}(z, t)$, $\hat{S}^{\dagger}(z, t)$ and $\hat{C}^{\dagger}(z, t)$ which create collective atomic excitations in $|p\rangle$, $|s\rangle$ and $|c\rangle$ respectively. In a one-dimensional approximation, these operators are governed by the following Heisenberg equations of motion,
\begin{align}
\label{eq: EIT equation 1}
\partial_t \hat{\mathcal{E}}(z, t) & = - c\partial_z \hat{\mathcal{E}}(z,t) + i G \hat{P}(z, t), \\
\partial_t \hat{P}(z, t) & = i G \hat{\mathcal{E}}(z, t) + i \Omega_s \hat{S}(z, t) - \gamma \hat{P}(z, t) + \hat{F}(z, t), \\
\label{eq: EIT equation 3}
\partial_t \hat{S}(z, t) & = i \Omega_s \hat{P}(z, t)  - i \int_0^L dz^{\prime} V_{z, z^{\prime}} \hat{C}^{\dagger}(z^{\prime}, t) \hat{C}(z^{\prime}, t) \hat{S}(z, t), \\
\label{eq: EIT equation 4}
\partial_t \hat{C}(z, t) & = - i \int_{0}^{L} dz^{\prime} V_{z, z^{\prime}} \hat{S}^{\dagger}(z^{\prime}) \hat{S}(z^{\prime}) \hat{C}(z, t).
\end{align}
Here, $c$ is the vaccum speed of light, $G=g\sqrt{\rho_a}$ is the collectively enhanced coupling of the $|g\rangle-|p\rangle$ transition (where $g$ is the single atom coupling and $\rho_a$ is the homogenous atomic density), $\Omega_s$ is the Rabi frequency of the classical crontrol field driving the $|p\rangle-|s\rangle$ Rydberg transition, and $\gamma$ is the decay rate of the intermediate state $|p\rangle$.  We assume low-intensity source and gate fields such the $|s\rangle - |s\rangle$ and $|c\rangle - |c\rangle$ interactions can be neglected.  The operator $\hat{F}(z, t)$ describes Langevin noise associated with the decay of the intermediate state \cite{Scully1997}.

Considering a system of $n_g$ stored gate excitions and $n_s$ incident source photons, we introduce $|\Psi_{n_g, n_s}\rangle$ as the initial state. In the Heisenberg picture, this can be constructed explicitly as,
\begin{equation}
\label{eq: state}
|\Psi_{n_g, n_s}\rangle = \frac{1}{\sqrt{n_g! n_s!}} \left[ \frac{1}{\sqrt{c}} \int_{-\infty}^{\infty} dz h(-z/c) \hat{\mathcal{E}}^{\dagger}(z,0) \right]^{n_s} \times  \left[ \int_{0}^{L}d\vec{z}_{n_g} \mathcal{C}(\vec{z}_{n_g})\hat{C}^{\dagger}(z_1,0) \dots \hat{C}^{\dagger}(z_{n_g},0) \right] |0\rangle,
\end{equation}
where $h(t)$ is the temporal mode of the incident (uncorrelated) source field, and $\mathcal{C}(\vec{z}_{n_g})$ is the initial spatial mode of the stored gate excitations where $\vec{z}_{n_g}\equiv z_1, z_2, \cdots, z_{n_g}$ denotes the vector of gate excitation coordinates. To determine the scattering-induced spin wave decoherence, it is necessary to consider the density matrix dynamics of the stored gate excitations. For this, we first define the operator $\hat{\rho}(\vec{x}_{n_g}, \vec{y}_{n_g}, t)$
\begin{equation}
\hat{\rho}(\vec{x}_{n_g}, \vec{y}_{n_g}, t) = \prod_{i=1}^{n_g} \hat{C}^{\dagger}(x_i, t) \prod_{i=1}^{n_g} \hat{C}(y_i, t).
\end{equation}
This can then be used in conjuction with Eq. (\ref{eq: state}) to define the elements of the stored spin wave density matrix as 
\begin{equation}
\rho_{n_s}(\vec{x}_{n_g}, \vec{y}_{n_g}, t) = \langle \Psi_{n_g, n_s}| \hat{\rho}(\vec{x}_{n_g}, \vec{y}_{n_g}, t) | \Psi_{n_g, n_s}\rangle,
\end{equation}
which characterises the spatial coherence between different configurations $\vec{x}_{n_g}$ and $\vec{y}_{n_g}$ of the stored gate excitations in response to scattering $n_s$ source photons. To evaluate the time dynamics of $\rho_{n_s}(\vec{x}_{n_g}, \vec{y}_{n_g}, t)$, we begin with the equation of motion for the coherence operator,
\begin{equation}
\label{eq: density matrix operator equation of motion}
\partial_t \hat{\rho}(\vec{x}_{n_g}, \vec{y}_{n_g}, t) = i \int_{0}^{L} dz \left[ \sum_k V_{z, x_k} - \sum_k V_{z, y_k} \right]  \hat{S}^{\dagger}(z, t) \hat{\rho}(\vec{x}_{n_g}, \vec{y}_{n_g}, t) \hat{S}(z, t),
\end{equation}
which can be readily derived from Eq. (\ref{eq: EIT equation 4}). The solution to the spin wave operator $\hat{S}(z, t)$ will be a convolution of the form
\begin{equation}\label{eq: S solution}
\hat{S}(z, t) = \int_{-\infty}^{\infty} dt^{\prime} \hat{e}(z, t - t^{\prime}) \hat{\mathcal{E}}(0, t^{\prime}),
\end{equation}
where $\hat{e}(z, t)$ is an operator object which is intrinsically nonlinear in the stored spin wave density $\hat{C}^{\dagger}(z, t)\hat{C}(z, t)$. The general solution also includes terms propotional to $\hat{\mathcal{E}}(z, 0)$, $\hat{P}(z, 0)$, $\hat{S}(z, 0)$ and $\hat{F}(z, 0)$. However, since all our results only involve normally ordered expectation values, such terms give vanishing contributions for the initial state in Eq. (\ref{eq: state}) \cite{Gorshkov2013, Murray2016a}. With the definition for $\hat{S}(z, t)$ in Eq. (\ref{eq: S solution}), the equation of motion for $\rho_{n_g}(\vec{x}_{n_g}, \vec{y}_{n_g}, t)$ can then be written as,
\begin{equation}\label{eq: density matrix equation of motion}
\begin{split}
\partial_t \rho_{n_s}(\vec{x}_{n_g}, \vec{y}_{n_g}, t) = i \frac{n_s}{c} \int_{0}^{L} dz & \left[ \sum_k V_{z, x_k} - \sum_k V_{z, y_k} \right] \int_{-\infty}^{\infty} dt^{\prime}h^*(t^{\prime})\int_{-\infty}^{\infty} dt^{\prime \prime}h(t^{\prime \prime}) \\
&\times  \langle \Psi_{ng, n_s - 1} |\hat{e}^{\dagger}(z, t - t^{\prime}) \hat{\rho}(\vec{x}_{n_g}, \vec{y}_{n_g}, t) \hat{e}(z, t - t^{\prime \prime}) |\Psi_{ng, n_s - 1}\rangle,
\end{split}
\end{equation}
where we have used the property $\hat{\mathcal{E}}(0, t)|\Psi_{ng, n_s}\rangle = \hat{\mathcal{E}}(-ct, 0)|\Psi_{ng, n_s}\rangle = \sqrt{n_s/c}h(t)|\Psi_{ng, n_s - 1}\rangle$. In the limit where the source field is narrowband in relation to the EIT bandwidth, we can make the replacement $\hat{e}(z, t) = \hat{e}(z)\delta(t)$, where $\hat{e}(z)$ defines the static solution to $\hat{S}(z, t)$ as $\hat{S}(z, t\to\infty) = \hat{e}(z)\hat{\mathcal{E}}(0, t\to\infty)$. This can be obtained by solving Eqs. (\ref{eq: EIT equation 1} - \ref{eq: EIT equation 3}) in the steady state to yield 
\begin{equation}
\label{eq: e operator}
\hat{e}(z) = - \frac{G}{\Omega_s}\frac{1}{1 + i\int dz^{\prime}  \mathcal{V}_{z, z^{\prime}} \hat{C}^{\dagger}(z^{\prime}) \hat{C}(z^{\prime})} 
 \exp \left( \frac{d_b}{z_b} \int_0^z dz^{\prime} \left[ \frac{1}{1 + i\int dz^{\prime \prime}  \mathcal{V}_{z^{\prime}, z^{\prime \prime}} \hat{C}^{\dagger}(z^{\prime \prime}) \hat{C}(z^{\prime \prime})} - 1 \right] \right),
\end{equation}
where $\mathcal{V}_{z, z^{\prime}} = \gamma V_{z, z^{\prime}} / \Omega_s^2$ is the rescaled interaction potential, and $2d_b = 2G^2 z_b / c \gamma$ is the optical depth per blockade radius, where $z_b$ is defined according to $V_{z_b, 0} = \Omega_s^2 / \gamma$. Eq. (\ref{eq: density matrix equation of motion}) can then be written as,
\begin{equation}
\label{eq: density matrix equation of motion 2}
\partial_t \rho_{n_s}(\vec{x}_{n_g}, \vec{y}_{n_g}, t) = i \frac{n_s}{c} |h(t)|^2 \int_0^L dz \left[ \sum_k V_{z, x_k} - \sum_k V_{z, y_k} \right] \langle \Psi_{ng, n_s - 1} |\hat{e}^{\dagger}(z) \hat{\rho}(\vec{x}_{n_g}, \vec{y}_{n_g}, t) \hat{e}(z) |\Psi_{ng, n_s - 1}\rangle.
\end{equation}

To proceed, we note that since the operator $\hat{e}(z)$ is constructed from the local density operator $\hat{C}^{\dagger}(z)\hat{C}(z)$, it conserves the total number of gate excitations. As such, the state $|C(\vec{x}_{n_g})\rangle = \prod_{i=1}^{n_g}\hat{C}^{\dagger}(x_i)|0\rangle$ is an eigenstate of $\hat{e}^{\dagger}(z)$ with an eigenvalue $e^*(z, \vec{x}_{n_g})$ defined by $\hat{e}^{\dagger}(z)|C(\vec{x}_{n_g})\rangle = e^*(z, \vec{x}_{n_g}) |C(\vec{x}_{n_g})\rangle$, which can be readily derived from Eq. (\ref{eq: e operator}) as 
\begin{equation}
e(z, \vec{x}_{n_g}) = - \frac{G}{\Omega_s}\frac{1}{1 + i\sum_{k} \mathcal{V}_{z, x_k}} 
 \exp \left( \frac{d_b}{z_b} \int_0^z dz^{\prime} \left[ \frac{\sum_{k} \mathcal{V}_{z, x_k}}{i - \sum_{k} \mathcal{V}_{z, x_k} } \right] \right).
\end{equation}
Upon then redefining $\hat{\rho}(\vec{x}_{n_g}, \vec{y}_{n_g}) = |C(\vec{x}_{n_g})\rangle \langle C(\vec{y}_{n_g})|$, it follows that the equation of motion for $\rho_{n_s}(\vec{x}_{n_g}, \vec{y}_{n_g}, t)$ can be written as
\begin{equation}
\label{eq: density matrix equation of motion 2}
\partial_t \rho_{n_s}(\vec{x}_{n_g}, \vec{y}_{n_g}, t) =n_s \phi_{n_g}(\vec{x}_{n_g}, \vec{y}_{n_g}) \rho_{n_s - 1}(\vec{x}_{n_g}, \vec{y}_{n_g}, t),
\end{equation}
where 
\begin{equation}\label{eq: phi}
\phi_{n_g}(\vec{x}_{n_g}, \vec{y}_{n_g}) = i \frac{d_b}{z_b}\frac{\Omega_s^2}{G^2} \int_0^L dz \left[ \sum_k \mathcal{V}_{z, x_k} - \sum_k \mathcal{V}_{z, y_k} \right]
 e^*(z, \vec{x}_{n_g}) e(z, \vec{y}_{n_g}),
\end{equation}
which defines the expression in Eq. (3) of the main text as $\Phi_{n_g}(\vec{x}_{n_g}, \vec{y}_{n_g}) = 1 + \phi_{n_g}(\vec{x}_{n_g}, \vec{y}_{n_g})$. The system of equations for $\rho_{n_s}(\vec{x}_{n_g}, \vec{y}_{n_g}, t)$ goverened by Eq. (\ref{eq: density matrix equation of motion 2}) can then be solved recusively in $n_s$ to yield the final expression for the many-body density matrix given in Eq. (2) of the main text.

\section{Spin wave decoherence in the infinite $d_b$ limit}
\label{sec: S2}
Here, we will derive the simple expression for the many-body density matrix in the infinite $d_b$ limit. Upon spatially ordering all gate excitations, whereby $x_1, y_1$ are the coherence coordinates of the first excitation, $x_2, y_2$ are the coordinates of the second and so on, then the result for $\Phi_{n_g}(\vec{x}_{n_g}, \vec{y}_{n_g})$ can be approximated as
\begin{equation}\label{eq: Phi approximate 1 - supplement}
\Phi_{n_g}(\vec{x}_{n_g}, \vec{y}_{n_g}) \approx 1 + \phi_1(x_1, y_1) +  (1 - p_{<2}) \phi_1(x_2, y_2) +  (1 - p_{<3}) \phi_1(x_3, y_3) + \cdots + (1 - p_{<n_g}) \phi_1(x_{n_g}, y_{n_g}),
\end{equation}
where $\phi_1(x_{k}, y_{k})$ is given by Eq. (\ref{eq: phi}), and $p_{<k}$ is the probability that a given source photon scatters before it reaches the $k^{\rm th}$ excitation. Here, it is implicitly assumed that $p_{<k}$ is close to unity, and in the infinite $d_b$ limit, one can make the approximation $p_{<k} = 1$. In this case, $\Phi_{n_g}(\vec{x}_{n_g}, \vec{y}_{n_g}) \approx 1 + \phi_1(x_1, y_1) = \Phi_1(x_{\rm min}, y_{\rm min})$ as given by Eq. (4) of the main text.

\section{Approximate model of retrieval efficiency}
\label{sec: S3}
Here, we will derive the approximate model of retrieval efficiency presented in Eq. (5) of the main text. We start by considering a system of $n_g$ stored gate excitations, and $n_s$ photons in the incident source field.  We assume a dilute system of excitations, such that the contributions from configurations of excitations with overlapping blockde radii can be neglected. The storage of such configurations will anyways be suppressed due to self-blockade between gate photons. As a second simplification, we assume that the scattering induced localisation of one gate excitaiton does not affect the mode shape, and thus retrieval, of any other. Formally, this approximation can be implemented by assuming the gate photons are stored in non-overlapping modes, and we introduce $\rho_{0}^{(k)}(x_{k}, y_{k})$ as the initial single body density matrix of the $k^{\rm th}$ excitation.   With this simplification, the initial many-body density matrix is given by the pure (uncorrelated) state $\rho_{0}(\vec{x}_{n_g}, \vec{y}_{n_g}) = \rho_{0}^{(1)}(x_1, y_1) \rho_{0}^{(2)}(x_2, y_2) \cdots \rho_{0}^{(n_g)}(x_{n_g}, y_{n_g})$. The efficiency of retrieving the $k^{\rm th}$ excitation after source photon scattering can be calculated from its reduced density matrix $\rho_{n_s}^{(k)}(x, y)$, which can be calculated from the full many-body density matrix according to 
\begin{equation}
\rho_{n_s}^{(k)}(x, y) = n_g \int dr_1 \cdots dr_{k-1}dr_{k+1} \cdots dr_{n_g} \rho_{n_s}(r_1, \cdots, r_{k-1}, x, r_{k+1}, \cdots, r_{n_g}, r_1, \cdots, r_{k-1}, y, r_{k+1}, \cdots, r_{n_g}).
\end{equation}
Assuming that the medium is much longer than the stored spin wave mode, the explicit form of $\rho_{n_s}^{(k)}(x, y)$ is given by,
\begin{equation}\label{eq: Reuced density matrix}
\rho_{n_s}^{(k)}(x, y) = \left[1 + A^{k - 1} \phi(x, y) \right]^{n_s} \rho_{0}^{(k)}(x, y),
\end{equation}
where $1 - A$ is the scattering probability per gate excitation defined according to,
\begin{align}
A & = \exp \left( \frac{d_b}{z_b} \int_{-\infty}^{\infty} dz^{\prime} \left[\frac{\mathcal{V}_{z^{\prime}, 0}}{i - \mathcal{V}_{z^{\prime}, 0} } - \frac{ \mathcal{V}_{z^{\prime}, 0}}{i + \mathcal{V}_{z^{\prime}, 0} } \right] \right), \\
& = \exp\left( 2 d_b \text{Re}\left[ \frac{2\pi}{3} (-1)^{11/12} \right] \right), \\
& \approx \exp(- 4 d_b).
\end{align}
The retrieval efficiency of the $k^{\rm th}$ excitation is then calculated as $\eta_{k}(n_s) = \mathcal{R}\left[\rho_{n_s}^{(k)}(x, y)\right]$. Here, $\mathcal{R}$ is a generic linear function for determining the retrieval efficiency from any given one-body density matrix and pulse sequence, whose explicit form is detailed in Ref. \cite{Gorshkov2007}. To simplify the calculation of $\eta_{k}(n_s)$, we assume that the blockade radius is much smaller than the spatial extent of each spin wave mode. In this situation, photon scattering will practically cause complete localisation of a given stored gate excitation. The quantity $\phi(x, y)$ characterising this decoherence in Eq. (\ref{eq: Reuced density matrix}) can then be approximated by
\begin{equation}
\phi(x, y) = \left\{
\begin{array}{ll}
0 & \quad \text{if}~x = y \\
A - 1 & \quad \text{otherwise}
\end{array}
\right.
\end{equation}
However, since the retrieval efficiency is predominatly determined by the spin wave coherences, it suffices to neglect the narrow digonal feature in $\phi(x, y)$ when caluclatuing $\eta_{k}(n_s)$. Using the approximation $\phi(x, y) \approx A - 1$, the retrieval efficiency of the $k^{\rm th}$ excitation is then given by
\begin{equation}
\eta_{k}(n_s) = \left[1 - p(1 - p)^{k - 1} \right]^{n_s} \mathcal{R}\left[\rho_{0}^{(k)}(x, y)\right],
\end{equation}
where we have used the fact that the scattering probability per gate excitation is given by $p=1-A$. We can then calculate the total number of retrieved gate photons from the stored $n_g$-excitation Fock state after scattering $n_s$ source photons as
\begin{equation}
\bar{n}_g(n_g, n_s) = \sum_{k=1}^{n_g} \eta_{k}(n_s) = \eta_R \sum_{k=1}^{n_g} \left[1 - p(1 - p)^{k - 1} \right]^{n_s},
\end{equation}
where we have made use of the fact that the retrieval function is linear, and further assumed that the retrieval efficiency in the absence of photon scattering is the same for all gate excitations, i.e., $\mathcal{R}\left[\rho_{0}^{(k)}(x, y)\right] = \eta_R$. Finally, taking into account the coherent state nature of the involved fields, we can calculate the average number of retrieved gate photons by performing a coherent state average of $\bar{n}_g(n_g, n_s)$ over the number distribution of the gate and source fields, which ultimately yields
\begin{align}
\bar{\alpha}_g & = e^{-\alpha_g}  e^{-\alpha_s}\sum_{n_g=1}^{\infty} \sum_{n_s=0}^{\infty} \frac{(\alpha_g)^{n_g}}{n_g!}\frac{(\alpha_s)^{n_s}}{n_s!} \bar{n}_g(n_g, n_s), \\
& = \eta_R e^{-\alpha_g}  e^{-\alpha_s} \sum_{n_g=1}^{\infty} \sum_{n_s=0}^{\infty} \frac{(\alpha_g)^{n_g}}{n_g!}\frac{(\alpha_s)^{n_s}}{n_s!} \sum_{k=1}^{n_g} \left[1 - p(1 - p)^{k - 1} \right]^{n_s}, \\
& = \eta_R e^{-\alpha_g}  \sum_{n_g=1}^{\infty} \frac{(\alpha_g)^{n_g}}{n_g!} \sum_{k=1}^{n_g} \exp\left[-\alpha_s p(1 - p)^{k - 1} \right].
\end{align}
Finally, we can calculate the retrieval efficiency as the ratio of the number of retrieved gate photons with and without source field scattering,
\begin{equation}
\eta = \eta_R \frac{e^{-\alpha_g}}{\alpha_g}  \sum_{n_g=1}^{\infty} \frac{(\alpha_g)^{n_g}}{n_g!} \sum_{k=1}^{n_g} \exp\left[-\alpha_s p(1 - p)^{k - 1} \right],
\end{equation}
as given by Eq. (5) of the main text.

\section{Single photon subtraction via decoherence}
\label{sec: S4}
Here we will derive a simple estimate for the efficiency of single photon subtraction based on the described decoherence mechanism. For this, first consider the operation using Fock states of the incoming gate and source fields. Let $|n_g\rangle$ describe the gate field containing $n_g$ photons, and $|n_s\rangle$ describe the source field containing $n_s$ photons. Through the combination of gate storage, source field scattering and gate retrieval, a perfectly functioning single photon subtractor will achieve the mapping $|n_g\rangle \mapsto |n_g-1\rangle$. Taking into account a finite storage and retrieval efficiency due to linear losses, this photon subtraction can be achieved either from failed storage or failed retrieval, the latter of which is controlled via scattering induced decoherence. 

\begin{figure}[h!]
\begin{center}
\includegraphics[width=0.9\columnwidth]{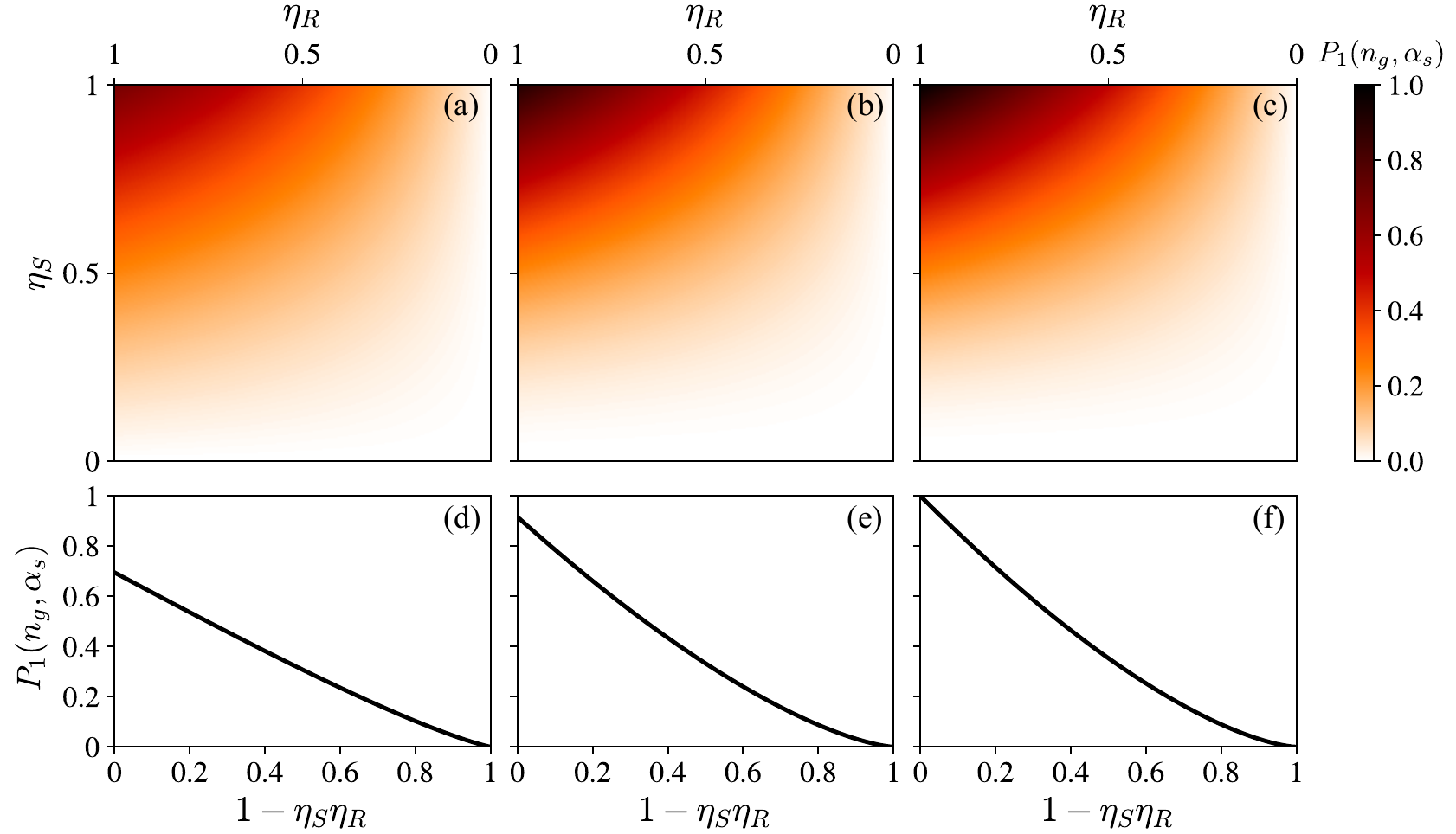}
\end{center}
\vspace{-4ex}
\caption{\label{fig: finite storage and retrieval}(a-c) The subtraction efficiency $P_1(n_g, \alpha_s)$ for a Fock state with $n_g=2$ incident gate photons is plotted as a function of the efficiency of storage, $\eta_S$, and retrieval, $\eta_R$. The blockaded optical depth is $d_b=0.5, 1$ and $5$ in (a), (b) and (c) respectively, and in each figure, we fix the coherent source field intensity to $\alpha_s^{(\rm opt)}$ which optimises $P_1(n_g, \alpha_s)$ for perfect storage and retrieval, $\eta_S=\eta_R=1$. $P_1(n_g, \alpha_s)$ is plotted as a function of the combined effieincy for storage and retrieval $\eta_S\eta_R$ (specifically for $\eta_S = \eta_R$) for $d_b=0.5, 1$ and $5$ in (d), (e) and (f) respectively.}
\end{figure}

To calculate the overall success probability for this to occur, let us first consider the storage losses. For this, we assume that storage is a linear process, and that each gate photon is stored with an probability $\eta_S$. The probability that all $n_g$ photons are succesfully stored, $P_0^{(S)}(n_g)$, and the probability that one fails to store, $P_1^{(S)}(n_g)$, are then each given by
\begin{align}
P_0^{(S)}(n_g) & = \eta_S^{n_g} \\
P_1^{(S)}(n_g) & = n_g (1 - \eta_S)\eta_S^{n_g - 1}
\end{align}
Assuming that $\bar{n}_g$ photons are stored, we then need to consider the subsequent decoherence dynamics from source field scattering. The probability $\text{p}_0(\bar{n}_g)$ that an incoming source photon fails to scatter from any of the $\bar{n}_g$ stored gate excitations is given by 
\begin{equation}
\text{p}_0(\bar{n}_g) = (1 - p)^{\bar{n}_g},
\end{equation}
and the probability $\text{p}_1(k_g)$ that a source photon scatters from the $k_g^{\rm th}$ excitation is given by,
\begin{equation}
\text{p}_1(k_g) = p(1 - p)^{k_g - 1},
\end{equation}
where $p$ is the scattering probability per gate excitation. The probability $P_0^{(D)}(\bar{n}_g, n_s)$ that none of the $n_s$ incoming source photons are scattered, such that $\bar{n}_g$ coherent excitations remain after the source field propagation, is then simply given by,
\begin{equation}
P_0^{(D)}(\bar{n}_g, n_s) = \left[ \text{p}_0(\bar{n}_g) \right]^{n_s}
\end{equation}
We then need to consider the probability that one gate photon is decohered after the source field scattering, which therefore leaves $\bar{n}_g - 1$ retrievable gate excitations. For this, the probability that $n_s$ incoming source photons decohere the $k_g^{\rm th}$ gate excitation only can then be considered as a sum of contributions: either all $n_s$ source photons scatter off the $k_g^{\rm th}$ excitation, or $n_s - 1$ source photons scatter off the $k_g^{\rm th}$ excitation while one is transmitted, or $n_s - 2$ source photons scatter off the $k_g^{\rm th}$ excitation while two are transmitted, and so on. The individual probabilities, $\mathcal{P}_{\bar{n}_g, n_s}^{(k_g, k_s)}$, that $k_s$ out of the $n_s$ incoming source photons scatter off the $k_g^{\rm th}$ gate excitation are then given by,
\begin{equation}
\mathcal{P}_{\bar{n}_g, n_s}^{(k_g, k_s)} = \binom{n_s}{k_s} \left[\text{p}_1(k_g) \right]^{k_s}\left[ \text{p}_0(\bar{n}_g) \right]^{n_s - k_s},
\end{equation}
where the binomial coefficient takes into account all the relevant scattering possibilities. The probability that at least one source photon scatters off the $k_g^{\rm th}$ gate excitation is then given by $ \sum_{k_s = 1}^{n_s}\mathcal{P}_{\bar{n}_g, n_s}^{(k_g, k_s)}$, such that the probability that only one gate excitation is left decohered after the passage of $n_s$ source photons is given by
\begin{align}
P_1^{(D)}(\bar{n}_g, n_s) & = \sum_{k_g = 1}^{\bar{n}_g} \sum_{k_s = 1}^{n_s} \mathcal{P}_{\bar{n}_g, n_s}^{(k_g, k_s)} \\
& = \sum_{k_g = 1}^{\bar{n}_g} \left\{ \sum_{k_s = 0}^{n_s} \binom{n_s}{k_s} \left[\text{p}_1(k_g) \right]^{k_s}\left[ \text{p}_0(\bar{n}_g) \right]^{n_s - k_s}  -  \left[ \text{p}_0(\bar{n}_g) \right]^{n_s} \right\} \\
& = \sum_{k_g = 1}^{\bar{n}_g} \left\{ \left[\text{p}_1(k_g) + \text{p}_0(\bar{n}_g) \right]^{n_s}  -  \left[ \text{p}_0(\bar{n}_g) \right]^{n_s} \right\}
\end{align}
We finally need to describe the linear retrieval losses, where we account for a finite retrieval probability of $\eta_R$ per gate excitation. Assuming that we are left with $\tilde{n}_g$ coherent gate excitations after source field scattering, the probability that all are successfully retrieved, $P_0^{(R)}(\tilde{n}_g)$, and the probability that one is lost during retrieval, $P_1^{(R)}(\tilde{n}_g)$, are then each given by
\begin{align}
P_0^{(R)}(\tilde{n}_g) & = \eta_R^{\tilde{n}_g} \\
P_1^{(R)}(\tilde{n}_g) & = \tilde{n}_g (1 - \eta_R)\eta_R^{\tilde{n}_g - 1}
\end{align}
The overall success probability for single photon subtraction $P_1(n_g, n_s)$ can than be evaluated by summing all contributions where exactly one photon is removed either during storage or retrieval, 
\begin{equation}
\begin{split}
P_1(n_g, n_s) = ~ & P_1^{(S)}(n_g)P_0^{(D)}(n_g-1, n_s) P_0^{(R)}(n_g - 1) \\
+ & P_0^{(S)}(n_g)P_1^{(D)}(n_g, n_s) P_0^{(R)}(n_g - 1) \\
+ & P_0^{(S)}(n_g)P_0^{(D)}(n_g, n_s) P_1^{(R)}(n_g) 
\end{split}
\end{equation}
The first line corresponds to single photon loss during storage, followed by perfect retrieval of all remaining stored excitations. The second line corresponds to successful storage of all gate photons, while one is removed upon retrieval due to scattering induced decoherence. Finally, the third line corresponds to successful storage of all gate photons, while one is removed upon retrieval due to linear losses. We can then use this result to obtain the success probability $P_1(n_g, \alpha_s)$ for single photon subtraction using a coherent source field containing an average number of photons $\alpha_s$ by performing a coherent state average of $P_1(n_g, n_s)$ over the number distribution of the source field,
\begin{equation}
P_1(n_g, \alpha_s) = e^{-\alpha_s} \sum_{n_s=0}^{\infty} \frac{\alpha_s^{n_s}}{n_s!}P_1(n_g, n_s)
\end{equation}
which is valid for $n_g>0$. At this point, we can examine the effects of imperfect storage and retrieval. For a given $n_g$, we can find the source field intensity $\alpha_s^{(\rm opt)}$ that optimises $P_1(n_g, \alpha_s)$ under conditions of perfect storage and retrieval, $\eta_S=\eta_R=1$. Considering a two-photon Fock state, we plot $P_1(n_g, \alpha_s^{(\rm opt)})$ against $\eta_S$ and $\eta_R$ in Fig. \ref{fig: finite storage and retrieval}, and further examine its scaling with the combined effieiency for storage and retrieval, $\eta_S\eta_R$. 

Finally, considering a coherent state of the gate field, we can define the averaged single photon subtraction efficiency $\mathcal{F}$ defined in Eq. 6 of the main text by performing a coherent state average over the number distribution of the gate field,
\begin{equation}
{\mathcal{F}} = e^{-\alpha_g} \left[ 1 + \sum_{n_g = 1}^{\infty} \frac{(\alpha_g)^{n_g}}{n_g!} P_1(n_g, \alpha_s) \right].
\end{equation}
Note that we implicitly set $P_1(n_g=0, \alpha_s)=1$, which assumes the subtraction is perfect for the vacuum component of the gate field. By optimisng $\mathcal{F}$ with respect to $\alpha_s$ for a given $\alpha_g$, we obtain the blue curve in Fig. 4 of the main text (where we consider perfect storage and retrieval efficiency).

\section{Single photon subtraction via saturable absorption}
\label{sec: S5}
Here, we will discuss the subtraction efficiency of the single photon absorber using a free-space Rydberg superatom, as recently demonstrated in \cite{Tresp2016}. The general mechanism in this case relies on saturating the absorption of an optically thick ensemble via Rydberg blockade. Here, engineered dephasing on the Rydberg state with a rate $\Gamma$ is used to achieve incoherent photon storage with a probability $p$. By working with a medium that is shorter than the blockade volume, the produced single Rydberg excitation then prevents any further photon absorption. For a large single photon detuning, the remaining off-resonant two-level medium is largely transparent to all subsequent photons, which scatter with a small residual probability $\tilde{p}$. Efficient single photon absorption with this mechanism then requires a large absorption probability $p$, while simultaneous minimising the residual photon losses.

This scheme is realised by coupling the quantised gate field to the low-lying excited state $|p\rangle$ with a large single photon detuning $\Delta$. A continuously applied control field then couples $|p\rangle$ to the Rydberg state $|s\rangle$ on two-photon resonance with a Rabi frequency $\Omega$. As before, one can introduce the operator $\hat{\mathcal{E}}^{\dagger}(z, t)$ to describe the creation of a gate photon, and introduce $\hat{P}^{\dagger}(z, t)$ and $\hat{S}^{\dagger}(z, t)$ to describe the creation of collective atomic excitations in $|p\rangle$ and $|s\rangle$. For a single incoming photon, the system dynamics are characterised by the following equations,
\begin{align}
\partial_t \hat{\mathcal{E}}(z, t) & = - c\partial_z \hat{\mathcal{E}}(z,t) + i G \hat{P}(z, t), \\
\partial_t \hat{P}(z, t) & = i G \hat{\mathcal{E}}(z, t) + i \Omega \hat{S}(z, t) -[ i\Delta +\gamma ]\hat{P}(z, t), \\
\partial_t \hat{S}(z, t) & = i \Omega \hat{P}(z, t) - \Gamma \hat{S}(z, t).
\end{align}
Here, Langevin noise can be neglected for the reasons outlined in Sec. \ref{sec: S1}. To zeroth order in the photon bandwidth, this system of equations reduces to a single propagation equation for $\hat{\mathcal{E}}(z)$ as
\begin{equation}
\partial_z \hat{\mathcal{E}}(z) = -\frac{1}{l_{\rm abs}} \frac{1}{\frac{\Gamma_{\rm EIT}}{\Gamma} + 1 + i \frac{\Delta}{\gamma}} \hat{\mathcal{E}}(z)
\end{equation}
where $\Gamma_{\rm EIT} = \Omega^2/\gamma$ is the resonant EIT bandwidth. For a medium of length $z_b$, the transmitted photon operator can be solved as
\begin{equation}
\hat{\mathcal{E}}(z_b) =  \exp\left[ - d_b \frac{1}{\frac{\Gamma_{\rm EIT}}{\Gamma} + 1 + i \frac{\Delta}{\gamma}}  \right]\hat{\mathcal{E}}(0) = \sqrt{1 - p} e^{i \theta} \hat{\mathcal{E}}(0),
\end{equation}
where $\theta$ is the phase of the transmitted field, and $p$ is the absorption probability, the latter of which is given by
\begin{equation}
p = 1 - \exp\left[ - 2 d_b \frac{1 + \frac{\Gamma_{\rm EIT}}{\Gamma}}{\left( 1 + \frac{\Gamma_{\rm EIT}}{\Gamma}\right)^2  + \left( \frac{\Delta}{\gamma} \right)^2 } \right].
\end{equation}
The residual (dissipative) scattering probability of the blockaded ensemble after photon absorption can then be straightforwardly obtained from the above expression by setting $\Gamma_{\rm EIT} = 0$ to give
\begin{equation}
\tilde{p} = 1 - \exp\left[ - 2 d_b \frac{1}{1 + \left(\frac{\Delta}{\gamma}\right)^2}  \right].
\end{equation}

To analyse the subtraction efficiency, first consider a Fock state of the incoming gate field containing $n_g$ photons. Treating the photons sequentially, the probability that exactly one is absorbed into the medium, whilst all others are transmitted can be calculated as 
\begin{equation}
P_1(n_g) = \sum_{k = 1}^{n_g} p(1 - p)^{k - 1} (1 - \tilde{p})^{n_g - k}.
\end{equation}
The subtraction efficiency for a coherent state with an average number of $\alpha_g$ photons is then given by
\begin{equation}
\label{eq: absorber subtraction efficiency}
\mathcal{F} = e^{-\alpha_g}\left[1 + \sum_{n_g = 1}^{\infty} \frac{(\alpha_g)^{n_g}}{n_g!} P_1(n_g)\right].
\end{equation}
For a given $d_b$ and $\alpha_g$, the optimal subtraction efficiency can be determined from Eq. (\ref{eq: absorber subtraction efficiency}) by maximising $\mathcal{F}$ with respect to $\Delta/\gamma$ and $\Gamma_{\rm EIT}/\Gamma$ to obtain the red dashed curve in Fig. 4 of the main text. Here, the additional constraint $\Gamma_{\rm EIT}/\Gamma \gg 1$ is imposed to ensure that incoherent photon absorption dominates over the dissipative scattering. 

\twocolumngrid
\bibliographystyle{apsrev4-1}
\bibliography{references}

\end{document}